%% file: main.tex
\documentclass[11pt]{article}

\usepackage[preprint]{acl}

\usepackage{times}
\usepackage{latexsym}

\usepackage[T1]{fontenc}

\usepackage{microtype}
\usepackage{etoc}
\usepackage{inconsolata}

\usepackage{graphicx}

\usepackage{enumitem}
\usepackage{amsmath}

\usepackage[utf8]{inputenc}
\usepackage{devanagari}

\usepackage{booktabs}
\usepackage{multirow}

\usepackage{enumitem}
\setlist{leftmargin=3pt} 

\usepackage{tabularx}
\usepackage{array}
\usepackage{booktabs}
\usepackage{pifont}
\usepackage{wasysym}

\usepackage{subcaption}
\usepackage{array}
\newcolumntype{C}[1]{>{\centering\arraybackslash}p{#1}}

\usepackage{colortbl}
\newcommand{\revision}[1]{{\color{black}#1}}
\usepackage[dvipsnames]{xcolor}
\usepackage{soul}
\definecolor{lightblue}{RGB}{173,216,230}
\title{
\makebox[\textwidth][c]{%
\raisebox{-0.5\height}{\includegraphics[height=1.6cm]{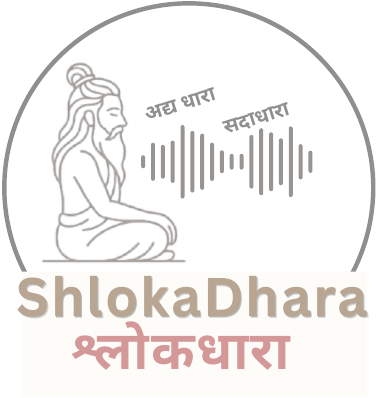}}%
\hspace{0.8em}%
\parbox[c]{0.75\textwidth}{\centering
ShlokaDhara: A Multi-Speaker Sanskrit Speech Corpus for Learning Meter-Aware Prosody in Text-to-Recitation}
}
}

\author{
  \textbf{Harshal Kausadikar\textsuperscript{1}}, 
  \textbf{Onkar Kishor Susladkar\textsuperscript{2}}, 
  \textbf{Prathmesh Santosh Choudhari\textsuperscript{3}}, \\
  \textbf{Vedangi Deepak Deshpande\textsuperscript{4}}, 
  \textbf{Ismini Lourentzou\textsuperscript{2}}, 
  \textbf{Sparsh Mittal\textsuperscript{5}} \\[0.1cm]
  \textsuperscript{1}\,Independent Researcher \texttt{harshalkausadikar@gmail.com} \\
  \textsuperscript{2}\,University of Illinois at Urbana-Champaign \texttt{\{onkarks2,lourent2\}@illinois.edu} \\
  \textsuperscript{3}\,University of Florida \texttt{ps.choudhari@ufl.edu} \\
  \textsuperscript{4}\,University of Minnesota - Twin Cities \texttt{deshp183@umn.edu} \\
  \textsuperscript{5}\,IIT Roorkee \texttt{sparsh.mittal@ece.iitr.ac.in}
}

\begin{document}
\maketitle
\input{00abstract}

\input{10intro}
\input{20relatedwork}

\input{30dataset}

\input{80results}

\input{95conclusion}
   
\bibliography{custom}
\renewcommand\thefigure{A.\arabic{figure}} 
\setcounter{figure}{0}

\renewcommand\thetable{A.\arabic{table}} 
\setcounter{table}{0}

\appendix

\input{100appendix}

\end{document}

%% file: 00abstract.tex
\begin{abstract}
Sanskrit verse recitation is governed by \textit{Chandas}, a strict prosodic framework defined by rhythmic \textit{Laghu} (light) and \textit{Guru} (heavy) syllables. Because existing Sanskrit speech datasets prioritize ASR or conversational TTS, they lack the metrical fidelity required for authentic verse synthesis. We introduce \textbf{\textit{ShlokaDhara}}, a 43-hour, multi-speaker Sanskrit speech corpus capturing diverse meters, recitation styles, and melodic variations. We specifically focus on \textit{Text-to-Recite}, a synthesis task explicitly requiring both prosodic accuracy and metrical correctness. \revision{Benchmarking state-of-the-art TTS architectures via an automated UTMOS (neural MOS) predictor demonstrates that XTTS-v2 achieves the highest perceptual quality. We further validate the XTTS-v2 generated samples using two paradigms: subjective assessments by 16 Sanskrit experts and novel empirical metrics, specifically, syllable-level \textit{Laghu-Guru} duration analysis to objectively quantify structural metrical awareness. Ultimately, this work establishes a robust dataset and evaluation benchmark for prosody-aware computational Sanskrit recitation.}
Generated Samples: \url{https://shlokadhara.vercel.app/}
\end{abstract}

%% file: 10intro.tex
\section{Introduction}
Sanskrit, often characterized as a ``formal language'' in natural guise, presents a unique paradigm
for computational linguistics due to its near-algorithmic structural precision. Revered as \textit{Devabh\={a}\d{s}\={a}} (the language of the gods), it reveals a rigorous
mathematical architecture in its poetry and possesses a Context-Free Grammar~\cite{ingerman1967panini,kadvany2016paṇini}. This structural precision extends to its prosody, where phonetics, rhythm, grammar and poetic structure are closely integrated. 

A substantial portion of the Sanskrit knowledge systems, including philosophy, mathematics,
astronomy, medicine, grammar, ethics, and devotional literature, was composed in metrical verse \cite{pingree1981jyotihsastra,  sarma2009sanskrit}.
This ``poem-first'' tradition enabled memorization, oral transmission
and preservation of knowledge across generations. Sanskrit literature is therefore predominantly
governed by rhythmic templates known as \textit{Chandas}. These meters are defined by systematic
sequences of \textit{Laghu} (light/short) and \textit{Guru} (heavy/long) syllables, which encode
the temporal structure of recitation. 
Sanskrit has more than 1,300 \textit{Chandas}, serving as the prosodic framework through which rhythm and metric structures are preserved.

Recitation is central to this tradition because Sanskrit knowledge was historically preserved not
only through manuscripts, but also through disciplined vocal transmission \cite{mair1991sanskrit}.
Even in modern education,
children learn foundational concepts through poems and songs because rhythm
improves retention. 
Classical Sanskrit works such as \textit{L\={i}l\={a}vat\={i}} expressed even mathematical ideas, including quadratic problems, in poetic form to support learning through chanting \cite{berg2001multiplication}. 
Sanskrit verse recitation is relevant for computational modeling, where speech synthesis must account for rhythmic and metrical characteristics in addition to pronunciation.\looseness-1

\begin{figure}[t!]
  \centering
  \includegraphics[width=0.7\columnwidth]{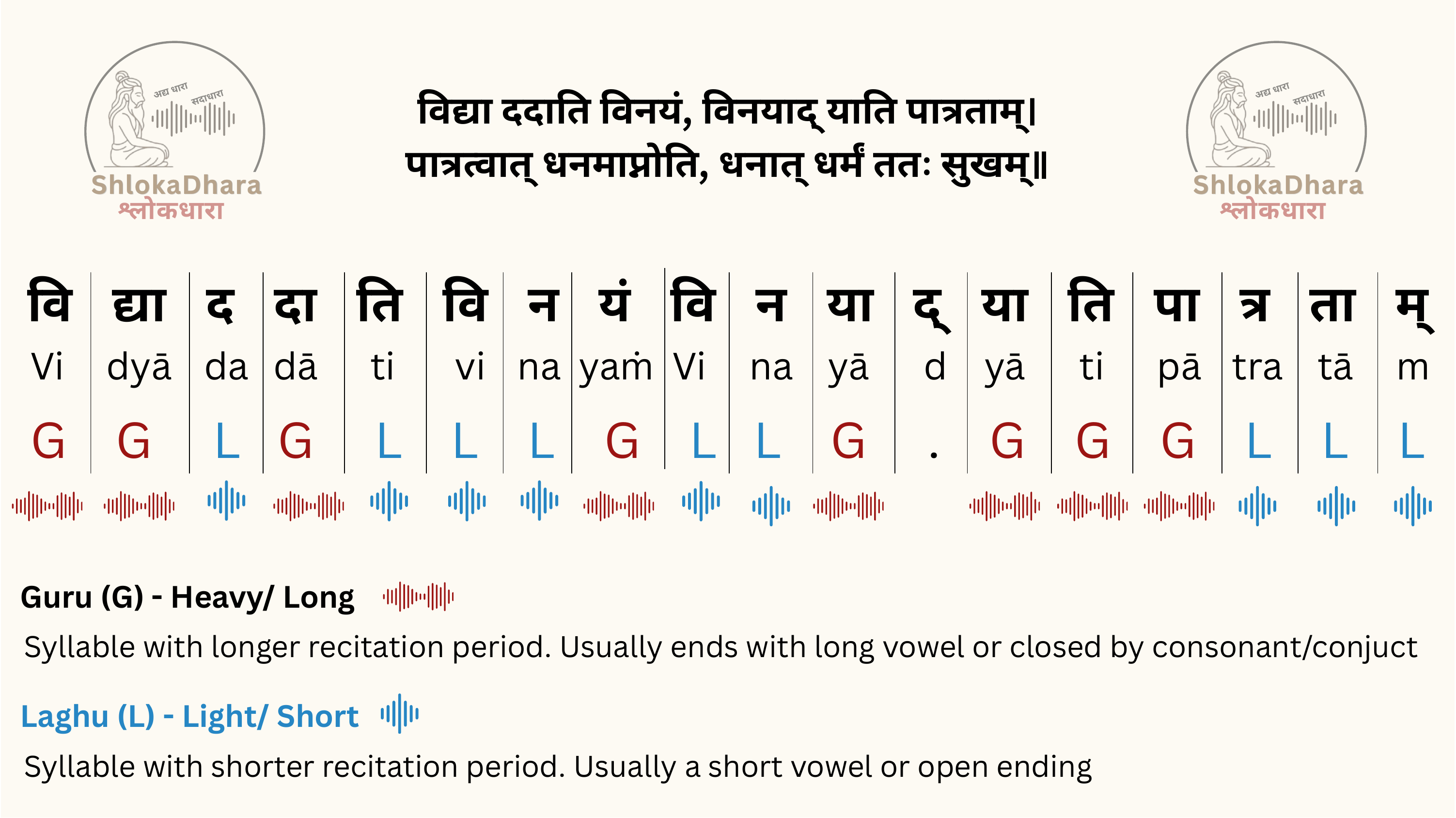}
  \caption{Syllable-level \textit{\textcolor{SkyBlue}{Laghu}} (L, light) and \textit{\textcolor{purple!70}{Guru}} (G, heavy) annotation for
    the \'{s}loka \textit{Vidy\={a} dad\={a}ti vinaya\d{m}, vinay\={a}d y\={a}ti p\={a}trat\={a}m} (``Knowledge gives humility; from humility comes worthiness''). }
  \label{fig:shloka}
\end{figure}

\noindent{\textbf{The Evolution of Prosodic Constraints:}}
Vedic Sanskrit prosody incorporated the \textit{Svara} system. Meters such as
\textit{G\={a}yatr\={i}}, \textit{Tri\d{s}\d{t}ubh}, and \textit{Jagat\={i}} were defined not
just by syllable count, but by a precise three-tone pitch system---\textit{Ud\={a}tta} (high),
\textit{Anud\={a}tta} (low), and \textit{Svarita} (falling)---which functioned as phonemic
markers. Later in Classical Sanskrit (\textit{Laukika}), prosodic organization shifted primarily toward quantitative meter, in which syllables are categorized as \textit{Laghu} (light/short) or \textit{Guru} (heavy/long) based on syllable weight and the time expected for their pronunciation. 
Figure~\ref{fig:shloka} illustrates this \textit{Laghu}-\textit{Guru} analysis for the
well-known \'{s}loka \textit{Vidy\={a} dad\={a}ti vinaya\d{m}}. (``Knowledge gives humility; from humility comes worthiness'')
 
Despite this structural rigidity, a significant computational gap exists: one \textit{Chanda} can
be realized through multiple acoustic melodies (tunes). However, contemporary neural
Text-to-Speech (TTS) systems frequently fail to maintain the metrical compliance required for these
rhythmic texts~\cite{jagadeeshan2026chandomitra}. 

To bridge this gap, we need a dataset where the same \textit{Chanda} is recorded across different
tunes by diverse speakers. This provides data for studying how a common metrical structure is realized across different melodic renditions and speakers. While large-scale speech corpora and expressive TTS datasets are available
for major Western languages such as English, French, and German, equivalent resources for Sanskrit
remain extremely limited. Existing Indian-language speech datasets primarily focus on conversational
speech, news reading, or ASR, where metrical consistency and
prosodic structure are not primary objectives. 

In this paper, we address this \textbf{Prosodic Bottleneck} by introducing a novel, high-fidelity
\textit{ShlokaDhara} dataset consisting of 43 hours of recorded Sanskrit \textit{Stotra}s, with the following
contributions:
\begin{itemize}[itemsep=0.5ex, parsep=0pt, topsep=-2.3pt, leftmargin=0.2cm]
   \item[\eighthnote] \textbf{Multi-Speaker Sanskrit Verse Text-to-Recite Dataset:}  
    We introduce a large-scale Sanskrit verse recitation dataset consisting of 19 speakers (11 male and 8 female) reciting verses across multiple \textit{Chanda}s. The dataset captures multiple melodic variations for the same metrical template. This enables the study and modeling of latent prosodic invariants across tunes and speakers.

  \item[\twonotes] \textbf{Tackling the Text-to-Recite Task:} 
    Motivated by the fundamentally metrical and prosodic nature of Sanskrit verses, this work addresses the \textit{Text-to-Recite} task. Unlike conventional TTS, which focuses on natural speech generation, the proposed task emphasizes preservation of metrical structure, rhythmic consistency, and recitation dynamics essential to Sanskrit verse rendering.

  \item[\eighthnote] \textbf{Benchmarking State-of-the-Art TTS Models for Sanskrit Verse Recitation:}   
    We benchmark seven state-of-the-art speech synthesis models (Section \ref{sec:results}) on the proposed Text-to-Recite task. Experimental evaluation demonstrates that our dataset can support high-quality Sanskrit speech synthesis across different modeling paradigms.
    
 \item[\twonotes] 
 \textbf{Objective Metrics for Metrical Analysis:} Standard speech synthesis is judged by how natural it sounds, but Sanskrit poetry must also follow strict timing rules. To verify this, we introduce a precise, syllable-level evaluation (Section~\ref{sec:metricalawareness}). By explicitly measuring the duration contrast between \textit{Laghu} (short) and \textit{Guru} (long) syllables, we provide a concrete benchmark to prove a model has actually learned the poetic meter. We will publicly release the dataset and source code.
\end{itemize}
 

%% file: 20relatedwork.tex
\section{Related Work}
\noindent\textbf{Sanskrit-Specific Automatic Speech Recognition (ASR) Datasets:} 
The \textit{Vāksañcayaḥ} corpus \citep{adiga2021automatic} provides 78 hours of diverse speech suitable for general acoustic modeling. 
Datasets such as IndicVoices \citep{javed2024indicvoices}, Kathbath \citep{javed2023indicsuperb}, 
and Shrutilipi \cite{bhogale2023effectiveness} scale to thousands of hours and capture  real-world conversational diversity by mining spontaneous speech and broadcast audio. However, by optimizing for sheer volume and conversational naturalness, these large-scale resources inherently lack the structured prosodic and metrical annotations required for classical verse synthesis.

For specialized religious domains, the Vedavani dataset \citep{kumar2025vedavani} captures 54 hours of Vedic chanting. While rhythmically rich, its reliance on single-source group chants limits its utility for TTS development. Furthermore, the pitch-based tonal accents of Vedic chanting differ fundamentally from the quantitative metrical structure of the Classical Sanskrit verses targeted in our work.

\noindent\textbf{Expressive Speech Datasets and TTS Models:} 
Compared to ASR, resources for Sanskrit speech synthesis remain sparse and prioritize general expressivity over metrical structure. On the data front, the Rasa dataset \citep{varadhan2024rasa} captures emotional variety, while corpora like IndicVoices-R \citep{sankar2024indicvoices} and IndicTTS \cite{baby2016resources} provide broad, multi-speaker Sanskrit speech (Table~\ref{tab:dataset_comparison}). On the modeling front, early systems \citep{debnath2020low, hear2read2024sanskrit} demonstrated viability using limited single-speaker transfer learning. Recent architectures like Indic-Speak \citep{indicspeak2026}, Indic-Parler-TTS \citep{sankar2025rasmalai}, and Bhashini-IITM TTS \cite{iit_madras_fastspeech2_2025} have since expanded to multi-speaker and expressive generation. However, while these combined advancements improve conversational naturalness, they implicitly treat prosody as a stochastic acoustic feature, lacking the deterministic focus required for rhythmic poetry.

\revision{Vāgdhenu \citep{prathosh2026vagdhenu}, a metrically-aware \textit{śloka}-to-chant synthesis technique, pairs a flow-matching backbone with a vṛtta-aware reference selection mechanism.
Crucially, in their model, learned text-side prosody conditioning proved architecturally inert, establishing that reference-driven acoustic conditioning is currently the most viable mechanism for ensuring strict metrical fidelity. However, Vāgdhenu's reference-based approach is bottlenecked by its highly constrained dataset: 5.3 hours of single-speaker audio in two pacing styles.}

\begin{table}[h]
\centering
\begin{tabular}{l c c c c}
\hline
\textbf{Dataset} & \textbf{Language} & \textbf{Hrs} & \textbf{Speakers} & \textbf{Type} \\
\hline
Vāksañcayaḥ & Sanskrit & 78+ & 27 & ASR \\
Kathbath & Multiple & 191.8 & Multiple & ASR \\
IndicVoices & Multiple & 23K+ & 51K+ & Conversational \\
Shrutilipi & Multiple & 20 & Multiple & Broadcast \\
Vedavani & Sanskrit & 54 & Limited & Vedic Chant \\
Rasa & Multiple & 20/spk & 44 & Expressive TTS \\
IndicVoices-R & Multiple & 35.75 & 161 & TTS \\
IndicTTS & Sanskrit & 22 & 2 & TTS \\
Vāgdhenu & Sanskrit & ~5.3 & 1 & Chant TTS \\
\textbf{ShlokaDhara} & \textbf{Sanskrit} & \textbf{43+} & \textbf{19} & \textbf{Recitation} \\
\hline
\end{tabular}
\caption{Comparison of existing Sanskrit and Indic datasets with \textit{ShlokaDhara}.}
\label{tab:dataset_comparison}
\end{table}

%% file: 30dataset.tex
\section{ShlokaDhara Dataset}
\subsection{Motivation and Contribution}
Three critical limitations emerge from the current landscape. (1)  large-scale datasets like IndicVoices and \textit{Kathbath} prioritize diversity over structured linguistic constraints. (2) Sanskrit-specific corpora such as \textit{Vāksañcayaḥ}, \textit{Shrutilipi}, and \textit{Vedavani} are optimized for ASR accuracy rather than prosodic modeling. (3) expressive datasets like \textit{Rasa} focus on emotional variation without enforcing metrical structure.
\revision{Crucially, except for the Vāgdhenu Sanskrit Chant Corpus, no existing dataset supports the development of meter-aware TTS systems. However, Vāgdhenu lacks the extensive acoustic variety and broad melodic diversity required to generalize zero-shot text-to-recitation across arbitrary voices.}

To resolve these fundamental data limitations we introduce \textbf{\textit{ShlokaDhara}} as a large-scale, multi-speaker, multi-tune Sanskrit TTS dataset explicitly designed to decouple invariant metrical structures from variable melodic realizations. By capturing various \textit{Chandas} across diverse tunes and speakers while maintaining strict metrical fidelity, our corpus provides an essential foundation for building robust, metrically-aware speech synthesis systems.

\textit{ShlokaDhara} consists exclusively of recited Sanskrit verses comprising over 43 hours of audio across diverse sources following LJSpeech Dataset \cite{ljspeech17} format. The dataset captures various rhythmic forms of Sanskrit, often derived from classical poetic meters. It opens up new avenues for exploring the musical dimensions of Sanskrit, extending beyond religious chanting to include classical poetry, drama, literature, and more. 
The dataset possesses several key features that distinguish it from existing Sanskrit TTS/ASR datasets. Firstly, it focuses exclusively on Sanskrit verses, highlighting the importance of prosody-driven linguistic structure. Secondly, it captures multiple prosodic variations or melodic renditions of the same meter, emphasizing the flexible and non-rigid musical nature of Classical Sanskrit in contrast to Vedic Sanskrit.
Finally, it includes recitations from 19 speakers across more than 50 sources, covering verses composed in over 20 different meters.

\subsection{Dataset Collection Pipeline}
Building a high-quality TTS dataset for Sanskrit is challenging due to its 
low-resource nature and the need for clean, high-fidelity recordings with 
consistent speaker characteristics. Our focus on verse-based content further limits the data availability, as preserving meter and recitation style is essential. Most publicly available sources are extemporaneous and do not follow the prosodic structure required for verse-based synthesis. For example, Sanskrit news broadcasts from All India Radio (AIR) are largely spontaneous and do not 
reflect the controlled rhythm and intonation of recited texts. Additionally, 
available recordings often suffer from background noise, inconsistent recording conditions, and a lack of speaker-wise separation. 

\begin{figure}[ht]
  \centering
  \includegraphics[width=0.7\columnwidth]{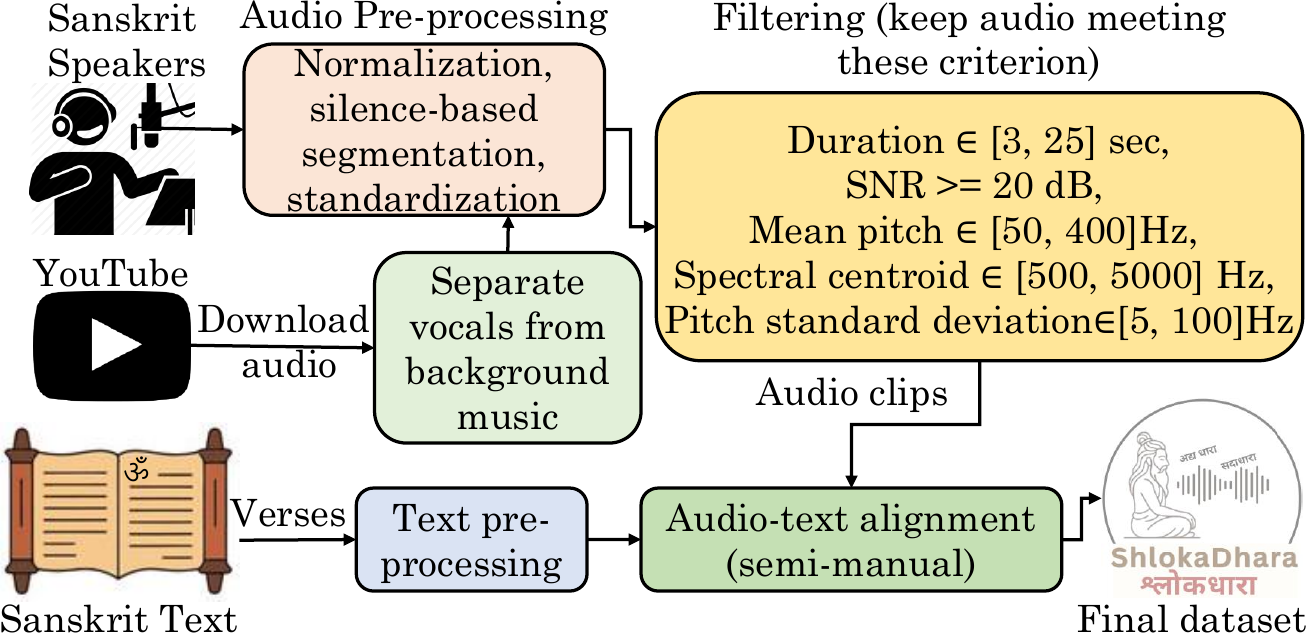}
  \caption{\textit{ShlokaDhara} Dataset Creation Pipeline}
  \label{fig:dataset_pipeline}
\end{figure}

Figure~\ref{fig:dataset_pipeline} shows the key steps in \textit{ShlokaDhara} dataset collection. We now discuss the three key steps in the dataset creation pipeline.

\subsubsection{Audio Processing}
\textbf{Step 1: Audio Collection} The Sanskrit audio recordings were obtained from two sources: YouTube and expert Sanskrit speakers. 
These audio samples were carefully curated to ensure high quality, including clear pronunciation, adherence to metrical rules of chanting, and consistency in melodic patterns throughout each recording. 
The recordings provided by Sanskrit speakers helped improve the coverage of underrepresented texts and, thus, enhance the richness of the dataset. 
These recordings,  made under  controlled conditions, capture variations in recitation styles, melodic interpretation and chanting pace .

\noindent \textbf{Step 2: Vocals Extraction} 
The YouTube audio files contained background music performed on traditional Indian instruments such as the Sitar, Flute, and Veena. We used the HTDemucs \citep{demucs} model to separate vocals from the background music to create a high-fidelity dataset.\\
\textbf{Step 3: Audio Preprocessing} 
We processed the audio files using the pydub library \citep{pydub}, which internally relies on FFmpeg \citep{ffmpeg} for audio decoding and encoding. Each audio file was first normalized to ensure consistent loudness across the dataset.
We then applied silence-based segmentation to identify speech regions using an energy-based voice activity detection approach. The segmentation process was not strictly rule-based; instead, the silence threshold and related parameters were adjusted iteratively based on the quality of the generated audio clips. In cases of
over-fragmentation or under-segmentation, these thresholds were manually tuned to obtain meaningful verse-level chunks.


For longer segments, a finer level of sub-segmentation was performed to ensure that each clip remained within a reasonable duration range suitable for training. This step also required multiple refinements, as different recordings exhibited varying levels of background noise, chanting styles, and pacing, which affected segmentation quality. Finally, all audio segments were standardized in terms of sampling rate and encoding format before being exported as individual WAV files.\\
\textbf{Step 4: Filtering} We apply a set of quality checks to filter out noisy or unreliable audio segments. Each segment is evaluated using the signal-to-noise ratio (SNR) along with spectral centroid as a timbral descriptor \citep{peeters2004large}. 
Using a YIN-based implementation \citep{de2002yin} from the \textit{librosa} Python library, statistical measures such as mean pitch and pitch standard deviation are computed to characterize pitch stability.
We removed audio segments that did not meet the following criteria:
SNR $\geq 20$ dB, spectral centroid  range $[500, 5000]$ Hz, mean pitch range $[50, 400]$ Hz, pitch standard deviation  range $[5, 100]$ Hz, and duration range of $[3, 25]$ seconds.

\subsubsection{Text Processing}
We collected text files of recited Sanskrit verses from reputable sources, viz., \textit{sanskritdocuments.org}  and \textit{stotranidhi.com}, to ensure consistency and accuracy.
The preprocessing pipeline involved removing extra spaces, non-Devanagari characters, numeric characters, and special symbols. We did not use YouTube transcripts since they are produced by general-purpose ASR models. These models do not accurately capture the pronunciation and phonetic distinctions of Sanskrit and hence, lead to a high amount of word- and phoneme-level errors.

\subsubsection{Audio-Text Alignment}
While audio cleaning and text normalization were largely automated, audio-text alignment could not be automated.  This step was the primary bottleneck, since variations in pacing and verse recitation make alignment more challenging.
Despite careful preprocessing, many audio segments did not align cleanly with verse boundaries, with some clips containing partial verses or abrupt splits, especially in cases of long compound words. 

To address this, we adopted a semi-manual approach to ensure high-quality (gold-standard) alignment. We first applied a duration-based heuristic to estimate the number of verses in each audio clip. Given an average duration per verse line $d_{\text{avg}}$, the number of lines for a clip of duration $d_i$ is computed as $n_i = \max \left(1, \left\lfloor \frac{d_i + 1}{d_{\text{avg}}} \right\rfloor \right)$. The offset accounts for pauses and variability in recitation speed, while the floor operation avoids over-allocation. Verses are then assigned sequentially without overlap, and manual corrections are applied when the heuristic fails due to irregular speaking patterns or segmentation errors.

\subsection{Dataset Composition and Statistics}
Our final dataset has 43+ hours of WAV-format audio at a 44.1 kHz sampling rate. It covers 19 speakers (8 female, 11 male) across 50+ text sources, with 30+ minutes of data from each speaker. 
Table~\ref{tab:dataset_stats} shows the dataset statistics. Refer Appendix~\ref{sec:appendix_speaker_wise_distribution} to understand the speaker wise contribution for dataset.

\begin{table}[ht]
\centering
\begin{tabular}{cc|cc}
\hline
\textbf{Metric} & \textbf{Value}  & \textbf{Metric} & \textbf{Value}\\
\hline
Duration (hours) & 43.24 & Avg.\ Duration (s) & 8.5 \\
Avg.\ Utterance Length & 6.02 tokens &\# Utterances & 18,322 \\
Avg.\ SNR (dB) & 32.03 & Spectral Centroid (Hz) & 2185.3 \\
Mean Pitch (Hz) & 210.97 & Pitch Std.\ Dev.\ (Hz) & 42.22 \\
Character Vocabulary & 64 & Word Vocabulary & 18218 \\
\hline
\end{tabular}
\caption{Summary statistics of the \textit{ShlokaDhara} dataset. All acoustic features are averaged over utterances.}
\label{tab:dataset_stats}
\end{table}

\noindent \textbf{Source Distribution:} \textit{Bhagavad Gita} recited by different speakers accounts for 75.7\% of the dataset (see Figure~\ref{fig:gita_speaker_distribution}).  
Despite a shared textual source, the recordings exhibit variation in recitation style, including differences in tune and tempo, resulting in melodically diverse samples. Top contributing sources other than \textit{Bhagavad Gita} are shown in Figure~\ref{fig:non_gita_sources}. (full list of sources is provided in Appendix~\ref{sec:appendix_source_wise_distribution}).
The following Chandas (in alphabetical order) are covered: Anushtup, Bhujangprayat, DrutVilambit, Harini, Indravajra, Indravansha, Malini, Mandakranta, Praharshini, Pushpitagra, Shalini, Shardulvikridit, Shikharini, Sragdhara, Totak, Trishtubh, Upajati,  Upendravajra,   Vaktram, Vanshastha and Vasantatilaka.

\begin{figure}[ht]
  \centering
  \includegraphics[width=0.6\columnwidth]{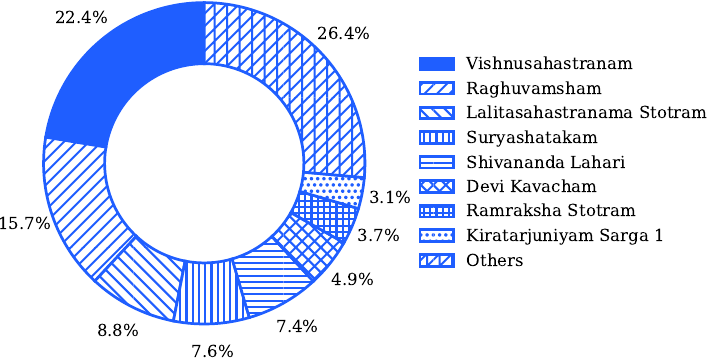}
  \caption{Top sources other than \textit{Bhagavad Gita}.}
  \label{fig:non_gita_sources}
\end{figure}

\begin{figure}[ht]
    \centering  
    \begin{subfigure}[b]{0.8\columnwidth}
        \centering
        \includegraphics[width=\linewidth]{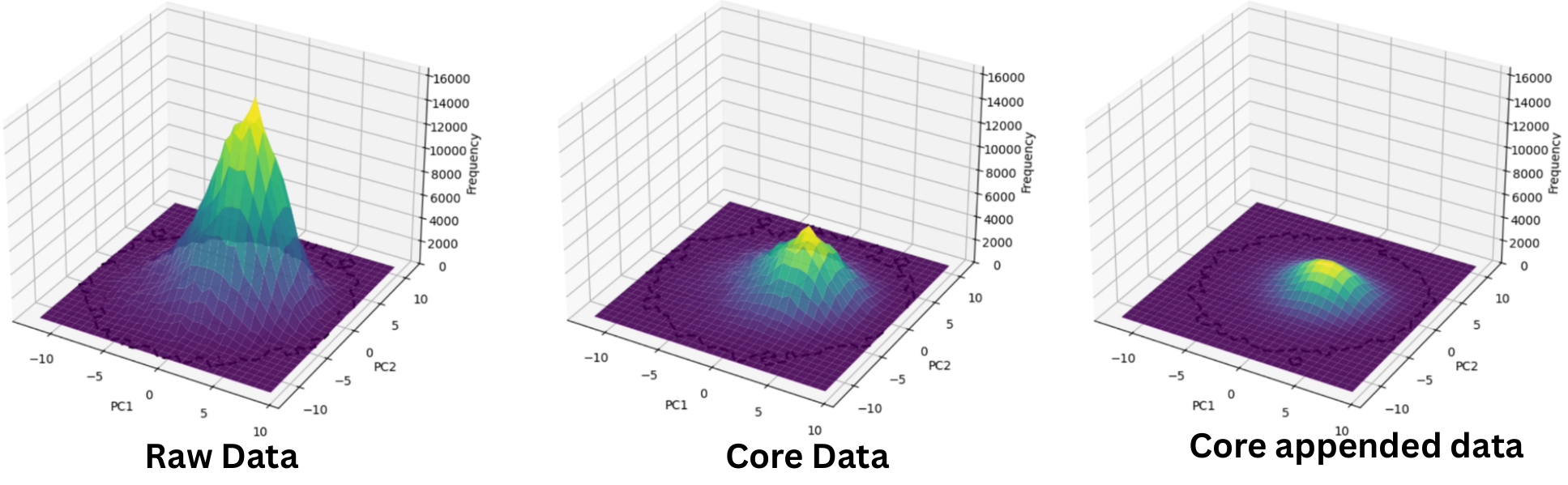}
        \caption{Semantic PCA with frequency and projection.}
        \label{fig:spca}
    \end{subfigure}
    
    \vspace{1em}
    
    \begin{subfigure}[b]{0.8\columnwidth}
        \centering
        \includegraphics[width=\linewidth]{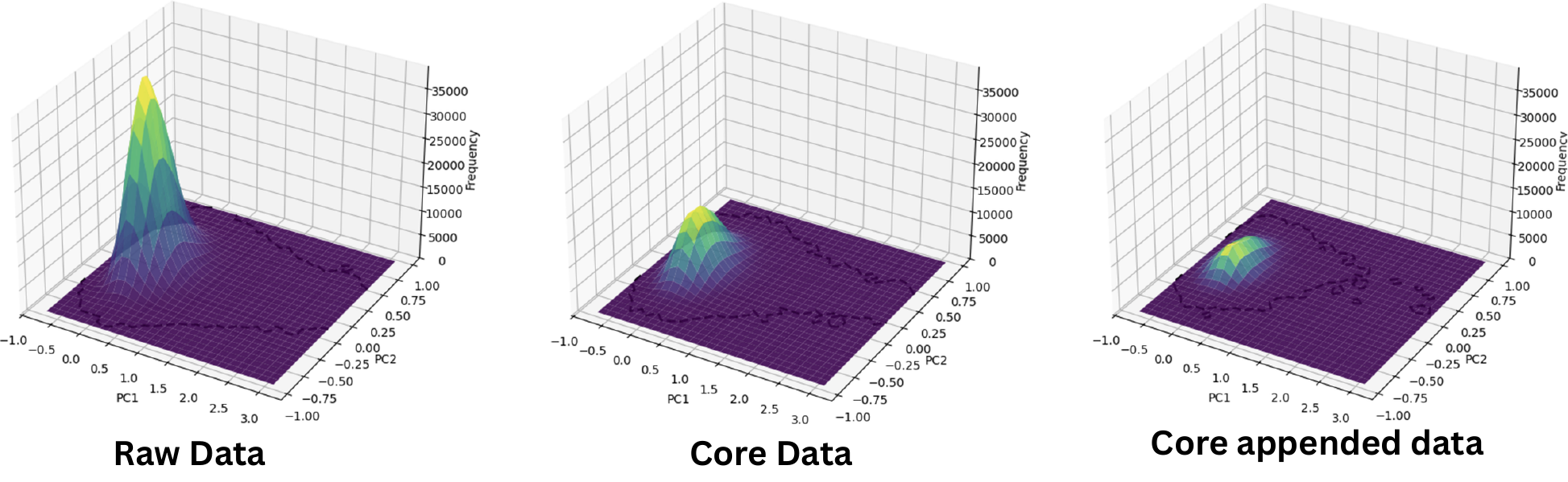}
        \caption{Acoustic PCA with frequency and projection.}
        \label{fig:apca}
    \end{subfigure}
    \caption{Semantic and acoustic PCA projections.}
    \label{fig:pca_combined}
\end{figure}

\noindent{\textbf{Analyzing semantic and acoustic diversity:}}
We analyze semantic and acoustic diversity using 768-dimensional embeddings from Sanskrit wav2vec~2.0~\cite{chadha2021vakyansh_sanskrit_wav2vec2} and Sanskrit-WavLM~\cite{sadhukhan2024automatic}, respectively. For each embedding space, we fit PCA on the raw dataset and project all dataset variants onto the same two principal components, enabling direct comparison (Figure \ref{fig:pca_combined}).
Coreset selection reduces concentration in densely populated regions. Following speaker diarization and UTMOS filtering, the retained data preserves over 95\% of the original semantic diversity and 85\% of the acoustic diversity. Duration appending largely preserves the core dataset's distribution in both embedding spaces.

\subsection{Comparing Speaker Melodic Diversity} 
To explicitly quantify and visualize the melodic diversity present in the \textit{ShlokaDhara} dataset, we analyzed the fundamental frequency ($F_0$) contours across different speakers reciting the same textual source (e.g., the \textit{Bhagavad Gita Chapter 2}). Because classical Sanskrit verses can be chanted in varying tunes and paces, isolating the melodic trajectory requires normalizing both pitch and time. The measurement pipeline was structured as follows:

\paragraph{1. Pitch Extraction and Normalization:}
For each audio segment, the fundamental frequency ($F_0$) was extracted using the probabilistic YIN (pYIN) algorithm. To eliminate absolute pitch differences between speakers (e.g., male versus female vocal registers) and focus strictly on the relative melodic excursions, we applied Z-score normalization to the valid $F_0$ values of each utterance:
\begin{equation}
    F_{0,\text{norm}} = \frac{F_0 - \mu_{F_0}}{\sigma_{F_0}}
\end{equation}
where $\mu_{F_0}$ and $\sigma_{F_0}$ are the mean and standard deviation of the utterance's fundamental frequency, respectively.

\paragraph{2. Time-Axis Interpolation:}
Because utterance durations naturally vary, both due to individual speaking rates and the differing lengths of the recited textual segments, the normalized $F_0$ contours were temporally aligned. Each contour was linearly interpolated onto a fixed percentage-based time axis ranging from 0\% to 100\% of the total utterance duration. 

\paragraph{3. Speaker-Level Aggregation:}
Finally, the time-normalized contours were averaged across all verses spoken by a given speaker to compute their average melodic profile.

\begin{figure}[h]
    \centering
    \includegraphics[width=0.75\columnwidth]{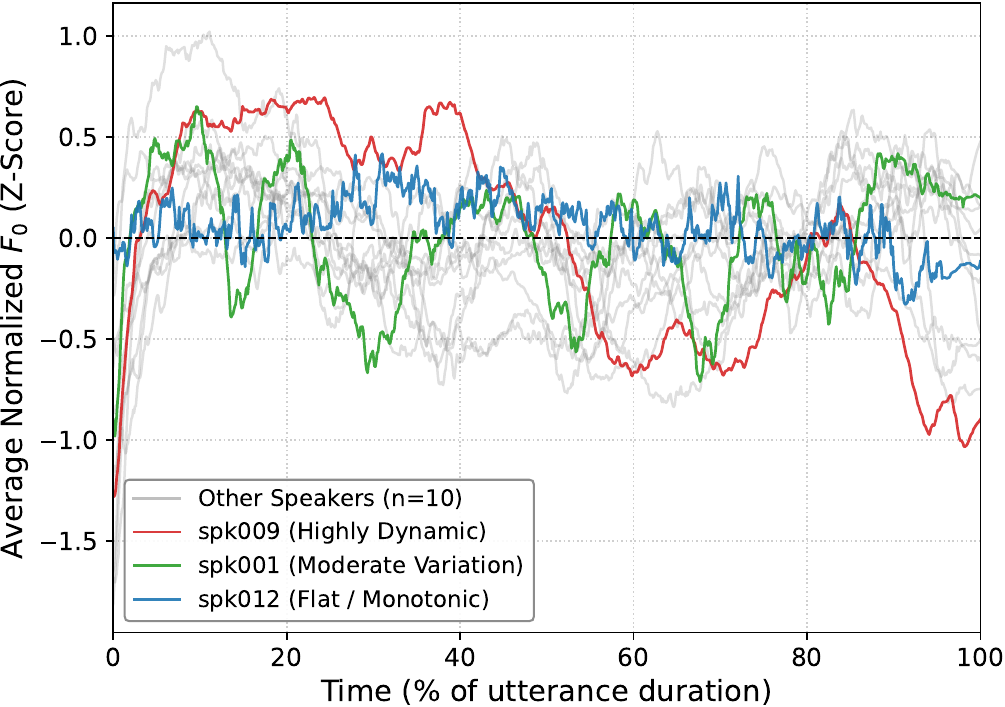}
    \caption{\revision{Average time-normalized $F_0$ contours (Z-scores) of 13 speakers reciting \textit{Bhagavad Gita Chapter 2}. Highlighting three distinct profiles (highly dynamic, moderate, flat/monotonic) against the remaining speakers demonstrates the dataset's broad melodic diversity.}}
    \label{fig:highlighted_f0_plot}
\end{figure}

As shown in Figure~\ref{fig:highlighted_f0_plot}, plotting these averaged trajectories reveals a wide spectrum of recitation styles. To effectively illustrate this variance without visual clutter, three representative speaker profiles are highlighted in the foreground: a highly dynamic contour (exhibiting wide pitch excursions), a moderate contour, and a flat/monotonic contour. The remaining speakers are plotted in the background (gray) to demonstrate the overall continuous variance space captured by the dataset.

%% file: 80results.tex
\section{Experiments and Results}
\subsection{Experimental Platform and Metrics}\label{sec:results}
\textbf{TTS Models:} 
We fine-tuned state-of-the-art TTS architectures, namely XTTS-v2 \citep{casanova2024xtts}, EnGenTTS \citep{susladkar2024bahasa}, Open-NaturalSpeechV2 \citep{shen2023naturalspeech}, Parler-TTS \citep{lyth2024natural}, Voxtral-TTS \citep{liu2025voxtral}, CLaM-TTS \citep{kim2024clam} and VITS \citep{kim2021conditional}, covering multiple architectural families and synthesis strategies, including flow-based models, diffusion-based models, language-model-based TTS systems, and neural vocoder-based end-to-end approaches.
We deliberately selected models that had already been trained on languages using the Devanagari script, since Sanskrit is commonly written in Devanagari. This choice allowed us to reuse existing text frontends and tokenizers,  avoiding the need to train a Sanskrit-specific tokenizer from scratch or manually extend the vocabulary with additional Sanskrit graphemes, ligatures, and phonetic units.

We fine-tuned each TTS model on the proposed \textit{ShlokaDhara} dataset using a unified experimental protocol wherever possible. All models were trained under identical optimization settings, while model-specific preprocessing components were retained to preserve compatibility with the original architectures. In particular, we used the corresponding text tokenizer or text frontend provided by each model, since different systems employ different tokenization strategies, such as character-level, subword-level, phoneme-level, or language-model-based tokenizers (refer to Appendix \ref{sec:appendix_exp}).

Our motivation is to assess whether the dataset can support high-quality Sanskrit verse recitation across different modeling paradigms, training objectives, acoustic modeling techniques, and speech generation mechanisms. This evaluation also helps determine how effectively pretrained multilingual or Devanagari-aware TTS models can adapt to Sanskrit pronunciation, prosody, phoneme-grapheme correspondence, and speaker characteristics when fine-tuned on our dataset.\looseness-1

\noindent\textbf{Train-Test Split:}
We split the dataset into \textit{seen} and \textit{unseen} subsets to test both in-domain performance and zero-shot generalization. The seen subset comprises $\sim$95\% of the corpus across 17 speakers. To benchmark in-domain synthesis, we applied an 80:20 train-test split to this data, maintaining a uniform speaker distribution to ensure balanced representation during training and evaluation.

Conversely, the unseen subset evaluates the model's zero-shot capabilities using the remaining 5\% of the data, which consists of utterances from two entirely held-out speakers. \revision{To strictly prevent data leakage, both these speakers and their corresponding source texts were excluded from the training phase. This disjoint test set provides a rigorous measure of the model's ability to generalize to unfamiliar voices, novel Sanskrit phrasing, and varying prosodic patterns}. This split allows us to analyze both the memorization-resistant synthesis quality and the robustness of each model under out-of-domain or less-familiar conditions.

\revision{
\textbf{Metric 1: UTMOS.} We evaluate the perceptual quality and naturalness of the generated speech using an Indic language-aware variant of UTMOS~\cite{saeki2022utmos}, a neural Mean Opinion Score (MOS) predictor that estimates human mean opinion scores directly from the speech waveform. 

Let $f_{\mathrm{UTMOS}}(\cdot)$ denote the pretrained UTMOS estimator.
Given $N$ synthesized utterances
$\mathcal{X}=\{x_i\}_{i=1}^{N}$, the reported UTMOS score is 
\begin{equation}
    \mathrm{UTMOS}
    =
    \frac{1}{N}
    \sum\nolimits_{i=1}^{N}
    f_{\mathrm{UTMOS}}(x_i),
    \label{eq:utmos}
\end{equation}
The model assigns each audio recording a quality score on a scale from 0 to 6;  higher scores indicate better predicted quality and naturalness. We compute UTMOS independently for each generated utterance and report the average over the corresponding evaluation split.

\textbf{Metric 2: Comparative MOS (CMOS).}
To measure the relative perceptual quality of synthesized speech with respect to the corresponding reference speech, we compute an automatic comparative MOS from the paired UTMOS scores. For the $i$-th utterance, the comparative score is defined as

\begin{equation}
    \mathrm{CMOS}_{i}
    =
    f_{\mathrm{UTMOS}}(x_i^{\mathrm{syn}})
    -
    f_{\mathrm{UTMOS}}(x_i^{\mathrm{ref}})
\end{equation}
 $x_i^{\mathrm{syn}}$ and $x_i^{\mathrm{ref}}$ denote the synthesized and reference speech, respectively. The final CMOS score is obtained by averaging over all $N$ utterances as $\mathrm{CMOS}
    =
    \frac{1}{N}
    \sum\nolimits_{i=1}^{N}    
    CMOS_i$.
A CMOS value closer to zero indicates a smaller perceptual-quality gap between the synthesized and reference speech. 
A  positive CMOS value
indicates that the synthesized speech has better perceived naturalness and overall speech quality than the reference.
The small negative magnitudes  reported in Table~\ref{tab:tts_quantitative_results} indicate that all evaluated models produce recitations that are only marginally degraded relative 
to the reference, highlighting the stronger resemblance with reference audio.}

\subsection{Quantitative Results}\label{sec:quantresults}
Table~\ref{tab:tts_quantitative_results} compares all fine-tuned TTS models on the proposed \textit{ShlokaDhara} dataset. 
Overall, XTTS-v2 achieves the highest UTMOS on both the seen set and unseen set, indicating that it produces the most natural and perceptually acceptable Sanskrit speech.
It also achieves the best CMOS score on the unseen set, suggesting stronger generalization to held-out Sanskrit utterances.

EnGenTTS obtains the second-highest UTMOS on the seen set ($3.87$), followed by CLaM-TTS ($3.78$) and Parler-TTS ($3.66$). These results suggest that modern neural TTS systems with stronger pretrained representations can effectively adapt to Sanskrit when fine-tuned on the proposed dataset. However, the performance gap between the seen and unseen sets indicates that generalization remains challenging, especially for models that are more sensitive to text distribution, pronunciation variation, or speaker-specific acoustic patterns.

\begin{table}[t!]
\centering
\begin{tabular}{lcccc}
\toprule
\multirow{2}{*}{\textbf{Model}} 
& \multicolumn{2}{c}{\textbf{Seen}} 
& \multicolumn{2}{c}{\textbf{Unseen}} \\
\cmidrule(lr){2-3} \cmidrule(lr){4-5}
& \textbf{UTMOS} $\uparrow$ 
& \textbf{CMOS} $\uparrow$
& \textbf{UTMOS} $\uparrow$ 
& \textbf{CMOS} $\uparrow$ \\
\midrule
VITS             & 3.12 & -0.0515 & 2.78 & -0.0611 \\
NaturalSpeech-V2 & 3.52 & -0.0503 & 2.82 & -0.0606 \\
EnGenTTS         & 3.87 & -0.0492 & 3.01 & -0.0598 \\
Parler-TTS       & 3.66 & -0.0442 & 2.56 & -0.0555 \\
Voxtral          & 3.52 & -0.0411 & 3.12 & -0.0562 \\
CLaM-TTS         & 3.78 & \textbf{-0.0372} & 2.89 & -0.0514 \\
XTTS-v2          & \textbf{4.01} & -0.0386 & \textbf{3.24} & \textbf{-0.0502} \\
\bottomrule
\end{tabular}
\caption{
Results of TTS models fine-tuned on the proposed \textit{ShlokaDhara} dataset (UTMOS range is 0 to 6)}
\label{tab:tts_quantitative_results}
\end{table}

CLaM-TTS and XTTS-v2 are among the two best models based on the CMOS score across seen/unseen datasets. Thus, their generated speech is perceptually close to the reference recordings under seen conditions, and preserves speech naturalness and acoustic fidelity under unseen conditions.
VITS shows lower scores due to its inability to adapt to the linguistic and phonetic complexity of Sanskrit compared with larger pretrained multilingual or language-model-based TTS systems.
Similarly, NaturalSpeech-V2 and Voxtral show moderate performance, while Parler-TTS exhibits a noticeable drop on the unseen set, showing weaker robustness under held-out evaluation conditions.

Overall, the results show that the proposed \textit{ShlokaDhara} dataset is suitable for fine-tuning a range of TTS architectures.
XTTS-v2's superior performance further suggests that multilingual pretrained models with robust text representations and acoustic priors are particularly effective for Sanskrit TTS adaptation. The gap between seen and unseen evaluation shows the need for further improvements in Sanskrit prosody modeling, pronunciation consistency, and generalization to diverse linguistic contexts.
\revision{For the experiments in Sections \ref{sec:humanevaluation} and \ref{sec:metricalawareness}, we exclusively evaluated XTTS-v2 generated samples due to its performance.}

\subsection{Human Evaluation of Text-To-Recite Task}\label{sec:humanevaluation}

\begin{figure}[ht]
  \centering
  \includegraphics[width=0.7\columnwidth]{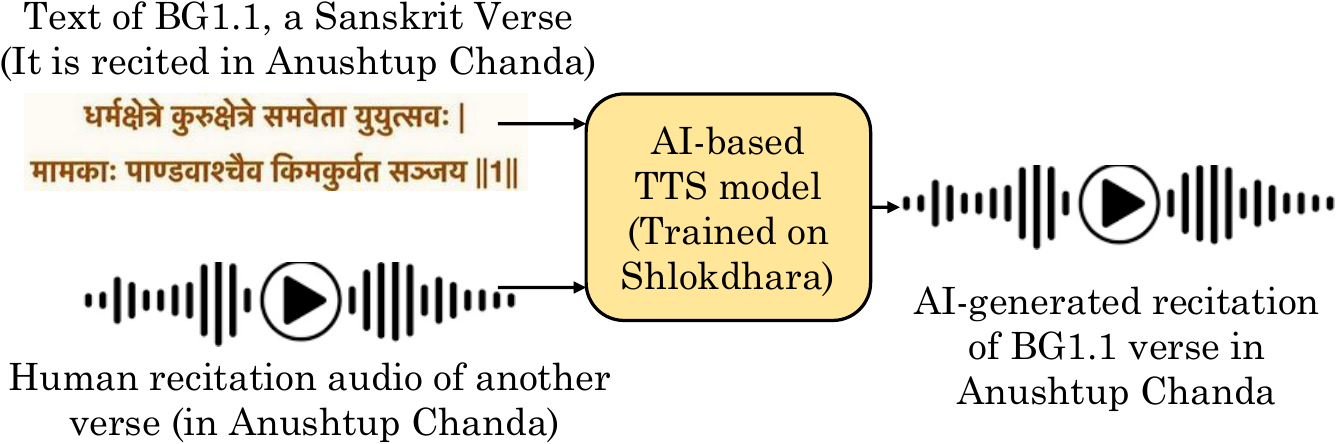}
  \caption{Evaluation of Text-to-Recite Task}
  \label{fig:ttr}
\end{figure}

During inference, the model takes a Sanskrit verse along with a reference recitation audio as input (Figure \ref{fig:ttr}). The reference audio is selected based on the \textit{Chanda} of the verse, allowing the model to capture both the speaker characteristics and the meter-specific recitation style while generating the output audio.  
The synthesized recitations are intended to preserve the rhythmic and prosodic structure associated with Sanskrit verse chanting.

We collected feedback from 16 Sanskrit experts on 20 synthesized audio clips comprising 10 male and 10 female voices across 9 different \textit{Chandas}. The reference audios used for synthesis were carefully selected to cover a range of linguistic and prosodic complexities, including challenging \textit{Sandhi} transitions, compound word formations, variations in tune, and differences in recitation speed. 
The evaluators assessed each synthesized audio sample using the criteria listed in Table~\ref{tab:text_to_recite_eval}. The evaluation criterion `Preservation of rhythmic and metrical structure' implicitly measures preservation of the \textit{Laghu–Guru} temporal structure associated with the target \textit{Chanda}. Each criterion was rated on a five-point Likert scale corresponding to \textit{Poor} (1), \textit{Fair} (2), \textit{Satisfactory} (3), \textit{Very Good} (4), and \textit{Excellent} (5).

The results indicate that XTTS-v2 achieved satisfactory performance across all assessment criteria, demonstrating the effectiveness of the metrically-aware \textit{ShlokaDhara} dataset for Sanskrit verse recitation synthesis.
\begin{table}[ht]
\centering
\begin{tabular}{cc}
\toprule
\textbf{Evaluation Criteria} & \textbf{Score (mean\revision{$\pm$95\% CI})} \\
\midrule
Clarity and naturalness of speech & $2.91 \revision{\pm 0.14}$ \\
Accuracy of pronunciation & $3.01 \revision{\pm 0.14}$ \\
Correctness of \textit{Sandhi} and word formation & $3.14 \revision{\pm 0.14}$ \\
Preservation of rhythmic and metrical structure & $2.92 \revision{\pm 0.14}$ \\
Overall quality of recitation & $2.92 \revision{\pm 0.13}$ \\
\bottomrule
\end{tabular}
\caption{Human evaluation results for the proposed Text-to-Recite task. \revision{(CI=confidence interval)}
}
\label{tab:text_to_recite_eval}
\end{table}

\revision{\subsection{Metrical Awareness Analysis}\label{sec:metricalawareness}

Evaluating the generated output is the best way to test whether \textit{ShlokaDhara} facilitates metrically-aware speech synthesis, since any model fine-tuned on our dataset should implicitly preserve Sanskrit metrical properties if the underlying training data successfully captures them. 
Designing an automated metric for Sanskrit prosody is beyond the scope of this work; we propose two practical empirical analyses to validate our hypothesis. 

\subsubsection{Controlled Laghu-to-Guru Sensitivity}
We seek to investigate whether the proposed \textit{ShlokaDhara} dataset enables a TTS model to capture the temporal significance of Sanskrit metrical structure. Since Guru syllables are inherently longer than Laghu syllables, increasing the metrical weight of a verse should generally increase its synthesized duration if the model has learned the underlying prosodic patterns.

For each evaluated verse, we construct a modified version by converting two or three Laghu syllables into their corresponding Guru forms within the first half of the verse while keeping the remaining text, reference audio, speaker characteristics, and synthesis parameters unchanged. Three substitutions are used whenever permitted by the text; otherwise, two substitutions are performed for shorter lines or verses with fewer eligible Laghu syllables. We intentionally modify multiple syllables because the temporal effect of a single Laghu-to-Guru substitution is typically too small to produce a reliable change in the overall utterance duration. Each modified verse is synthesized five times, and the average duration across all runs is used for analysis.

\begin{table}[ht]
\centering
\revision{
\begin{tabular}{lclc}
\toprule
Statistic & Value & Statistic & Value\\
\midrule
Evaluated verses & 17 & Inference runs per verse & 5\\
Verses with increased duration & 13 (76.5\%) & Mean unmodified duration & 11.95 s\\
Mean modified duration & 12.25 s & Mean duration increase & 0.30 s\\
\bottomrule
\end{tabular}
}
\caption{\revision{Utterance-level Laghu--Guru sensitivity results}}
\label{tab:lg_summary}
\end{table}

Table~\ref{tab:lg_summary} summarizes the results. 
Out of the 17 evaluated verses, thirteen (76.5\%) exhibit an increase in synthesized duration after the controlled Laghu-to-Guru modifications, indicating that the generated speech generally responds to an increase in metrical weight learned from the proposed dataset. 
With a mean of $\sim$2.1 substitutions per verse and an empirical Laghu-to-Guru contrast of $\Delta \approx 0.142$\,s (Table~\ref{tab:full_verse_lg_duration_ratios}), the predicted net lengthening is $2.1 \times 0.142 \approx 0.298$\,s, which matches the observed increase of $\sim$0.30\,s.

While the utterance-level analysis provides evidence of metrical sensitivity, the total utterance duration alone cannot directly verify whether the intended syllables are being realized with the correct temporal contrast.
During speech synthesis, the overall duration is influenced not only by the realization of individual syllables but also by prosodic factors, e.g., pause placement~\cite{panja2013critical}, phrase boundaries, and local speaking rate. Hence, the temporal increase introduced by the modified Guru syllables may be partially compensated elsewhere, 
preventing a consistent increase in the overall duration even when the modified syllables themselves are lengthened. For the remaining four verses (23.5\%), overall duration did not expand because the model applies prosodic rate compensation (isochrony), slightly compressing downstream syllables or intra-line pauses to preserve the cadence of the recited meter.
 
\subsubsection{Laghu–Guru Duration Analysis}
To more precisely assess metrical learning, we now quantify the temporal relationship between heavy (Guru) and light (Laghu) syllables in the 20 XTTS-v2 generated samples.
In classical Sanskrit prosody, a Guru syllable carries twice the moraic weight of a Laghu syllable, establishing a theoretical 2:1 duration ratio \citep{deo2007metrical}. 

Since automated phonetic aligners often struggle with the complex phrasing of synthesized Sanskrit audio, we ensured measurement precision by manually annotating boundaries.  We first extracted the ground-truth metrical sequence for each verse using Chandojnanam \citep{terdalkar2022chandojnanam}, and then manually annotated the syllable boundaries in Praat \citep{boersma2001praat}. These alignments allowed us to calculate the average syllable durations per verse (full breakdown in Appendix \ref{sec:appendix_LG_ratio_analysis}).
As shown in Figure~\ref{fig:lg_duration_ratio_analysis}, the average Guru-to-Laghu duration ratio across the annotated samples is 1.55:1. Although this deviates from the strict 2:1 mathematical ideal, it confirms the model successfully learns a distinct temporal contrast between syllable weights. Across quantitative metrical languages, empirical phonetics demonstrates that heavy syllables consistently compress to an acoustic ratio between 1.35:1 and 1.6:1 rather than strict 2:1 doubling \citep{hubbard1995toward, broselow1997syllable}. Although quantified cross-linguistically, this moraic timing framework directly parallels the Sanskrit Laghu-Guru system. Our observed 1.55:1 ratio directly reflects this phonetic reality.
In natural speech and by extension, highly expressive autoregressive models' rigid moraic boundaries are naturally softened by speaking rate, phrasing, and co-articulation. Thus, the observed 1.55 ratio shows strong fidelity to the underlying meter without forcing the synthesized recitation into an unnatural, robotic cadence.
}

%% file: 95conclusion.tex
\section{Conclusion}
We introduced \textbf{\textit{ShlokaDhara}}, a large-scale, multi-speaker Sanskrit speech corpus, and evaluated models on the \textit{Text-to-Recite} task to jointly preserve metrical structure and naturalness in verse synthesis. Benchmarking across diverse TTS architectures demonstrated the dataset's efficacy;  subjective human evaluations confirm the high perceptual quality of the generated recitations. \revision{Crucially, we extended standard evaluation paradigms by introducing syllable-level empirical metrics—specifically a \textit{Laghu-Guru} duration analysis. This objective assessment proved that models fine-tuned on our dataset successfully learn the strict temporal contrast between heavy and light syllables.} Ultimately, this work establishes a robust data and evaluation foundation for metrically-aware speech synthesis in highly structured languages, opening new directions for Sanskrit computational linguistics and prosody modeling.

\section{Limitations}
While \textit{ShlokaDhara} represents an initial step toward Sanskrit text-to-recitation modeling, two main limitations remain. First, the dataset's metrical distribution is heavily skewed toward the \textit{Anushtup} meter. While this naturally reflects its status as the most prevalent meter in classical Sanskrit literature (including foundational texts like the \textit{Bhagavad Gita}), this inherent imbalance means models may generalize less effectively to more complex, underrepresented \textit{Chandas}. \revision{Second, our fine-grained syllable-level \textit{Laghu-Guru} duration analysis currently relies on manual boundary annotation. Because standard automated aligners degrade significantly on the complex phrasing of stylized Sanskrit chant, manual intervention was strictly necessary to guarantee ground-truth precision. Future work will focus on expanding the corpus with diverse literary genres to balance the metrical distribution, and on developing specialized phonetic aligners tailored for stylized recitation to fully automate the evaluation pipeline.}

%% file: 100appendix.tex
\section*{Appendix}
This appendix provides supplementary materials, detailed experimental setups, and further analyses supporting the findings in the main text. Table \ref{tab:appendix_overview} outlines the contents of the following sections.

\begin{table}[ht]
\centering
\begin{tabular}{l l}
\hline
\textbf{Section} & \textbf{Description} \\
\hline
\ref{sec:appendix_speaker_wise_distribution}. Speaker-wise Duration Distribution & Details audio duration contributed by each speaker. \\
\ref{sec:appendix_source_wise_distribution}. Source-wise Duration Distribution & Details the textual source breakdown of the corpus. \\
\ref{sec:appendix_LG_ratio_analysis}. Laghu-Guru-Duration Ratio Analysis & Syllable duration analysis proving metrical awareness. \\
\ref{sec:appendix_exp}. Experimental details & Provides hyperparameters and setup for reproducibility. \\
\hline
\end{tabular}
\caption{Table of Contents of the Appendix Section}
\label{tab:appendix_overview}
\end{table}

\section{Speaker-wise Duration Distribution}
\label{sec:appendix_speaker_wise_distribution}

\begin{figure}[h]
  \centering
  \includegraphics[width=0.7\columnwidth]{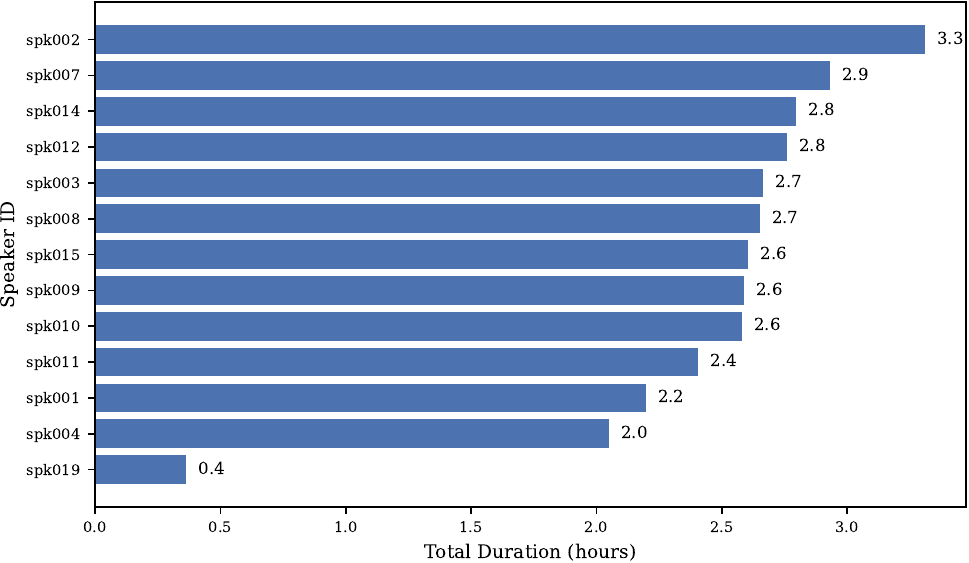}
  \caption{Speaker-wise distribution for Bhagavad Gita recited by 12 speakers, with some chapters from one speaker}
  \label{fig:gita_speaker_distribution}
\end{figure}

Table \ref{tab:speaker_duration} shows speaker-wise duration contribution in hours for the entire \textit{ShlokaDhara} dataset, sorted in descending order. Figure \ref{fig:gita_speaker_distribution} shows  Speaker-wise distribution for Bhagavad Gita recited by 12 speakers, with some chapters from one speaker. 

\begin{table}[h]
\centering
\setlength{\tabcolsep}{8pt}
\begin{tabular}{l r @{\hspace{1.2cm}} l r}
\toprule
\textbf{Speaker} & \textbf{Hours $\downarrow$} &
\textbf{Speaker} & \textbf{Hours $\downarrow$} \\
\midrule
spk012 & 3.42 & spk011 & 2.41 \\
spk014 & 3.35 & spk017 & 2.38 \\
spk002 & 3.31 & spk004 & 2.32 \\
spk015 & 3.10 & spk001 & 2.20 \\
spk007 & 2.93 & spk013 & 1.95 \\
spk003 & 2.66 & spk006 & 1.52 \\
spk008 & 2.65 & spk005 & 1.50 \\
spk009 & 2.59 & spk019 & 1.20 \\
spk010 & 2.58 & spk016 & 0.68 \\
       &      & spk018 & 0.50 \\
\bottomrule
\end{tabular}
\caption{Speaker-wise duration contribution in hours for complete dataset, sorted in descending order.}
\label{tab:speaker_duration}
\end{table}

\section{Source-wise Duration Distribution}
\label{sec:appendix_source_wise_distribution}
Table \ref{tab:source_distribution} shows the source-wise duration distribution (in hours), number of unique speakers, and associated chandas for each source for the entire \textit{ShlokaDhara} dataset.
\begin{table}[h]
\centering
\small
\begin{tabular}{l c c l}
\toprule
\textbf{Source} & \textbf{Hours $\downarrow$} & \textbf{Speakers} & \textbf{Chandas} \\
\midrule
Gita Chapter 2 & 3.29 & 13 & Anushtup, Upajati, Indravajra, Upendravajra \\
Gita Chapter 18 & 3.11 & 11 & Anushtup, Trishtubh \\
Gita Chapter 11 & 3.08 & 12 & Anushtup, Upajati, Indravajra, Upendravajra \\
Vishnusahastranam & 2.54 & 6 & Anushtup, Shalini, Vanshastha, Indravajra \\
Gita Chapter 6 & 2.08 & 12 & Anushtup \\
Gita Chapter 4 & 2.04 & 13 & Anushtup \\
Gita Chapter 1 & 2.01 & 12 & Anushtup \\
Gita Chapter 3 & 1.97 & 13 & Anushtup \\
Gita Chapter 10 & 1.85 & 12 & Anushtup \\
Gita Chapter 9 & 1.59 & 12 & Shalini, Anushtup \\
Gita Chapter 13 & 1.50 & 12 & Anushtup \\
Gita Chapter 7 & 1.38 & 12 & Anushtup \\
Gita Chapter 5 & 1.37 & 12 & Anushtup \\
Gita Chapter 8 & 1.36 & 12 & Anushtup, Upajati, Indravajra \\
Gita Chapter 14 & 1.25 & 12 & Anushtup \\
Gita Chapter 17 & 1.06 & 10 & Anushtup \\
Gita Chapter 16 & 1.06 & 11 & Anushtup \\
Lalitasahastranama Stotram & 1.00 & 2 & Anushtup, Trishtubh, Sragdhara \\
Gita Chapter 12 & 0.98 & 13 & Anushtup \\
Gita Chapter 15 & 0.93 & 11 & Anushtup, Upajati, Indravajra \\
Suryashatakam & 0.86 & 1 & Anushtup \\
Shivananda Lahari & 0.83 & 1 & Shikharini, Shardulvikridit \\
Raghuvamsham Sarga 2 & 0.83 & 2 & Upendravajra, Anushtup, Praharshini \\
Devi Kavacham & 0.56 & 2 & Anushtup \\
Raghuvamsham Sarga 1 & 0.44 & 1 & Anushtup \\
Ramraksha Stotram & 0.42 & 3 & Anushtup, Bhujangprayat, Shardulvikridit \\
Kiratarjuniyam Sarga 1 & 0.35 & 1 & Vanshasth, Malini \\
Amogh Shiv Kavacham & 0.29 & 1 & Anushtup, Vasantatilaka, Indravajra, Upendravajra \\
Raghuvamsham Sarga 12 & 0.26 & 1 & Upajati \\
Raghuvamsham Sarga 9 & 0.25 & 1 & DrutVilambit \\
Shri Mahalakshmi Suprabhatam & 0.23 & 1 & Vasantatilaka, Anushtup, Totak, Shardulvikridit \\
Aditya Hridayam & 0.21 & 2 & Anushtup \\
Shri Suktam & 0.19 & 2 & Anushtup, Bhujangprayat, Indravajra, Pushpitagra \\
Abhigyan Shakuntalam Act 4 & 0.18 & 1 & Shardulvikridit \\
Shri Mahalakshmi Stotram & 0.18 & 1 & Vaktram, Anushtup \\
Meghdutam & 0.15 & 1 & Mandakranta \\
Chanakyaneeti & 0.14 & 1 & Anushtup, Shardulvikridit, Bhujangprayat \\
Kanakadhara Stotram & 0.14 & 1 & Vasantatilaka \\
Siddha Kunjika Stotram & 0.12 & 2 & Anushtup, Harini \\
SB Canto1 Chapter 9 & 0.09 & 1 & Anushtup, Vanshastha, Pushpitagra \\
SB Canto1 Chapter 8 & 0.09 & 1 & Anushtup, Vanshastha, Vasantatilaka \\
SB Canto1 Chapter 7 & 0.09 & 1 & Upendravajra, Anushtup, Indravajra \\
SB Canto1 Chapter 5 & 0.09 & 1 & Indravajra, Vanshasth, Anushtup \\
Argala Stotram & 0.08 & 1 & Anushtup \\
SB Canto1 Chapter 3 & 0.08 & 1 & Vanshasth, Anushtup, Indravajra, Upendravajra \\
SB Canto1 Chapter 11 & 0.08 & 1 & Anushtup, Vanshastha, Vasantatilaka \\
SB Canto1 Chapter 6 & 0.07 & 1 & Anushtup, Indravansha \\
SB Canto1 Chapter 10 & 0.07 & 1 & Anushtup, Vanshastha, Pushpitagra \\
SB Canto1 Chapter 4 & 0.07 & 1 & Vasantatilaka, Anusthup, Vanshasth \\
SB Canto1 Chapter 2 & 0.06 & 1 & Vasantatilaka, Anushtup \\
Nirvana Shatakam & 0.06 & 2 & Bhujangprayat \\
Mahavir Ashtakam & 0.06 & 1 & Shikharini \\
Manisha Panchakam & 0.05 & 1 & Anushtup \\
Guru Ashtakam & 0.05 & 1 & Bhujangprayat \\
SB Canto1 Chapter 1 & 0.05 & 1 & Shardulvikridit, Drutvilambit, Anushtup \\
RSS Prayer & 0.02 & 1 & Bhujangprayat \\
\bottomrule
\end{tabular}
\caption{Source-wise duration distribution (in hours), number of unique speakers, and associated chandas for each source.}
\label{tab:source_distribution}
\end{table}
\clearpage

\revision{\section{Laghu-Guru-Duration Ratio Analysis}
We plot the empirical durations of Laghu and Guru syllables in Figure~\ref{fig:lg_duration_ratio_analysis}. The diagonal spread of the data points directly reflects recitation speed: points in the lower-left denote faster chanting, whereas those in the upper-right indicate a slower pace. 

\label{sec:appendix_LG_ratio_analysis}
\begin{figure}[h]
\centering
  \includegraphics[scale=0.7]{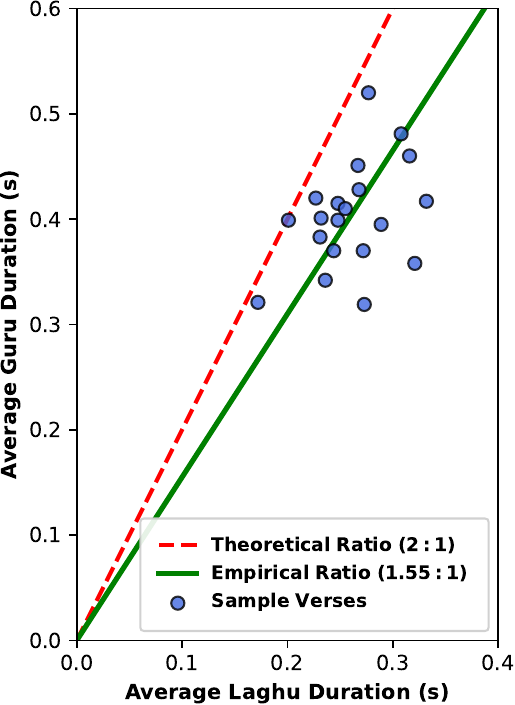}
  \caption{\revision{Laghu-Guru Duration Analysis Plot}}
  \label{fig:lg_duration_ratio_analysis}
\end{figure}

Complementing this visual summary, Table~\ref{tab:full_verse_lg_duration_ratios} details the verse-by-verse syllable counts and exact duration measurements for the generated audio samples, sequenced as presented on our project website: \url{https://shlokadhara.vercel.app/}.

\begin{table}[h]
\small
\centering
\revision{
\begin{tabular}{lccccc}
\toprule
\textbf{ID} & \textbf{\# Laghu} & \textbf{\# Guru} & \textbf{Laghu (s)} & \textbf{Guru (s)} & \textbf{Ratio} \\
\midrule
1 & 13 & 19 & 0.232 & 0.401 & 1.72 \\
2 & 12 & 20 & 0.321 & 0.358 & 1.12 \\
3 & 18 & 16 & 0.201 & 0.399 & 1.98 \\
4 & 18 & 16 & 0.172 & 0.321 & 1.87 \\
5 & 32 & 16 & 0.289 & 0.395 & 1.37 \\
6 & 16 & 8 & 0.272 & 0.370 & 1.36 \\
7 & 12 & 20 & 0.332 & 0.417 & 1.25 \\
8 & 12 & 20 & 0.227 & 0.420 & 1.85 \\
9 & 13 & 19 & 0.316 & 0.460 & 1.46 \\
10 & 8 & 15 & 0.231 & 0.383 & 1.66 \\
11 & 18 & 16 & 0.277 & 0.520 & 1.88 \\
12 & 6 & 16 & 0.308 & 0.481 & 1.56 \\
13 & 12 & 20 & 0.244 & 0.370 & 1.52 \\
14 & 15 & 19 & 0.267 & 0.451 & 1.69 \\
15 & 18 & 24 & 0.268 & 0.428 & 1.60 \\
16 & 24 & 15 & 0.273 & 0.319 & 1.17 \\
17 & 14 & 14 & 0.248 & 0.415 & 1.67 \\
18 & 14 & 14 & 0.236 & 0.342 & 1.45 \\
19 & 14 & 14 & 0.248 & 0.399 & 1.61 \\
20 & 19 & 25 & 0.255 & 0.410 & 1.61 \\
\midrule
\textbf{Avg} & \textbf{15.4} & \textbf{17.3} & \textbf{0.261} & \textbf{0.403} & \textbf{1.55} \\
\bottomrule
\end{tabular}
}
\caption{\revision{Verse-by-verse breakdown of syllable counts and empirical durations. ID: Sample ID, \# Laghu / \# Guru: Laghu/Guru Count, Laghu / Guru (s): Average Laghu/Guru Duration and Guru to Laghu Ratio. For averaging the count and duration, arithmetic mean has been used. For averaging the ratio (last column), geometric mean has been used.}}
\label{tab:full_verse_lg_duration_ratios}
\end{table}

The overall syllable composition leans slightly heavy. On average, verses contain a higher number of Guru syllables compared to Laghu syllables.
The data also highlights a massive variance in raw timing. The absolute durations for both syllable types fluctuate widely across the entire set, proving that the recitation tempo is highly dynamic rather than locked into a single fixed speed.}

\section{Experimental details}
\label{sec:appendix_exp}
For the acoustic input, all speech waveforms were resampled to a sampling rate of $22.05$ kHz. Each audio signal was represented as a one-dimensional temporal waveform, $x = \{x_t\}_{t=1}^{T}$, and $x_t \in [-1,1]$, where $x_t$ denotes the normalized amplitude value at time step $t$. Unless otherwise required by a specific model implementation, acoustic feature extraction parameters such as hop length, number of Mel bins, window length, and FFT size were kept identical to the default configuration of the respective pretrained model. This design choice minimizes architectural mismatch and allows each model to benefit from its original acoustic representation.

To maintain consistency during training, we used a batch size of $32$ with gradient accumulation over $4$ steps, resulting in an effective batch size of $B_{\mathrm{eff}} = 32 \times 4 = 128$.
All experiments were conducted on $8$ NVIDIA H100 GPUs, each with $80$ GB of memory. Each model was fine-tuned for approximately $2.5$K optimization steps. We used AdamW as the optimizer, with an initial learning rate of $1 \times 10^{-5}$. The learning rate was annealed using a cosine decay schedule throughout training. For AdamW, we used momentum parameters consistent with the model-specific default settings, with $\beta_2 = 0.99$ where applicable.

Overall, this setup was designed to isolate the effect of model architecture on Sanskrit TTS adaptation while keeping the training recipe, audio resolution, optimization strategy, and training duration as consistent as possible across all evaluated systems.